\documentclass[%
 reprint,
 amsmath,amssymb,
prl,
]{revtex4-2}

\usepackage{color}
\usepackage{rotate}
\usepackage{graphicx}
\usepackage{dcolumn}
\usepackage{bm}

\newcommand\Nu{\mbox{\textit{Nu}}}
\newcommand\Ra{\mbox{\textit{Ra}}}
\newcommand\Pran{\mbox{\textit{Pr}}}

\newcommand\Ro{\mbox{\textit{Ro}}}
\newcommand\Ek{\mbox{\textit{Ek}}}

\newcommand{\avg}[2]{\langle#1\rangle_{\mathrm{#2}}}

\begin{document}

\title{Boundary Zonal Flow in Rotating Turbulent Rayleigh--B{\'e}nard Convection}

\author{Xuan~Zhang$^{1}$}
\thanks{These two authors contributed equally.}
\author{Dennis~P.~M.~van~Gils$^{1,2}$}
\thanks{These two authors contributed equally.}
\author{Susanne~Horn$^{1,3,4}$}
\author{Marcel~Wedi$^{1}$}
\author{Lukas~Zwirner$^{1}$}
\author{Guenter~Ahlers$^{1,5}$}
\author{Robert~E.~Ecke$^{1,6}$}
\author{Stephan~Weiss$^{1,7}$}
\author{Eberhard~Bodenschatz$^{1,8,9}$}
\author{Olga~Shishkina$^{1}$}
\email{Olga.Shishkina@ds.mpg.de}
\affiliation{$^{1}$Max Planck Institute for Dynamics and Self-Organization, 37077 G\"ottingen, Germany }
\affiliation{$^{2}$Physics Fluids Group, J.M. Burgers Center for Fluid Dynamics, University of Twente, 7500 AE Enschede, The Netherlands}
\affiliation{$^{3}$Department of Earth, Planetary, and Space Sciences, UCLA, CA 90095, USA}
\affiliation{$^{4}$Centre for Fluid and Complex Systems, Coventry University, Coventry CV1 5FB, UK}
\affiliation{$^{5}$Department of Physics, University of California, Santa Barbara, CA 93106, USA}
\affiliation{$^{6}$Center for Nonlinear Studies, Los Alamos National Laboratory, Los Alamos, New Mexico 87545, USA}
\affiliation{$^{7}$Max Planck -- University of Twente Center for Complex Fluid Dynamics}
\affiliation{$^{8}$Institute for Nonlinear Dynamics, Georg-August-University G\"ottingen, 37073 G\"ottingen, Germany}
\affiliation{$^{9}$Laboratory of Atomic and Solid-State Physics and Sibley School of Mechanical and Aerospace Engineering, Cornell University, Ithaca, NY 14853, USA}

\date{\today}

\begin{abstract}
%
For rapidly rotating turbulent Rayleigh--B\'enard convection in a slender cylindrical cell,  experiments and direct numerical simulations reveal a boundary zonal flow (BZF) that replaces the classical large-scale circulation. The BZF is located near the vertical side wall and enables enhanced heat transport there.  {\color{black} Although the azimuthal velocity of the BZF is cyclonic (in the rotating frame), the temperature is an anticyclonic traveling wave of mode one whose signature is a bimodal temperature distribution near the radial boundary.  The BZF width is found to scale like $\Ra^{1/4}\Ek^{2/3}$ where the Ekman number $\Ek$ decreases with increasing rotation rate.}
\end{abstract}

\maketitle

\begin{figure*}
\includegraphics{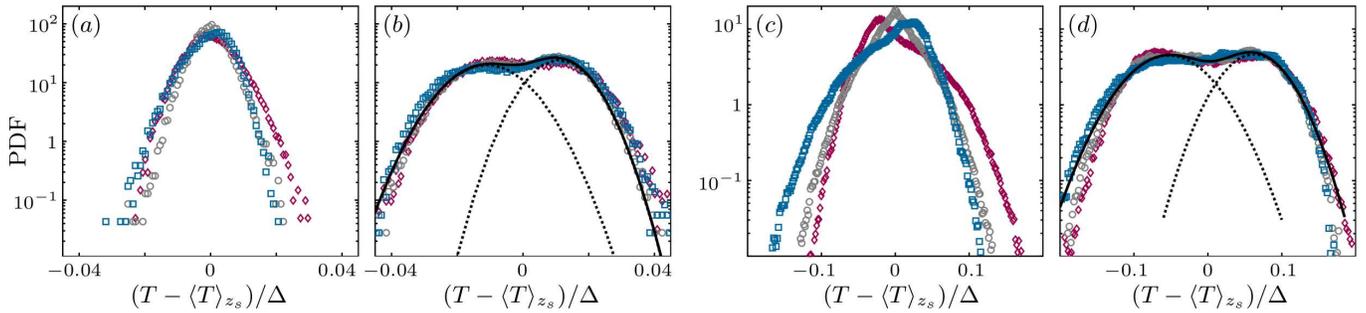}
\caption{Sidewall temperature PDFs, $r/R=1$, for $z/H=1/4$ (diamonds), $z/H=1/2$ (circles), and $z/H=3/4$ (squares), with $1/\Ro=0$  $(a,\,c)$ and 10 $(b,\,d)$. Experimental measurements with $\Ra=8\times10^{12}$ $(a,\,b)$ and DNS with $\Ra=10^{9}$ $(c,\,d)$, both with $\Pran=0.8$. Bimodal Gaussian distributions (solid lines), the sum of two normal distributions (dashed lines), are observed for rapid rotation $(b,\,d)$. 
$\langle {\cdot}\rangle_{z_s}$ denotes average in time and over all sensor positions at distance $z$ from the hot plate.
}
\label{PIC1}
\end{figure*}

\begin{figure}
\includegraphics{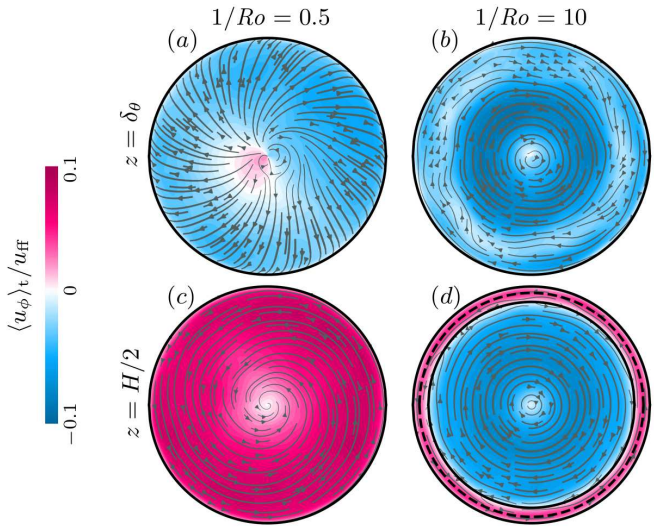}
\caption{
Horizontal cross-sections of time-averaged flow fields (DNS), visualized with streamlines (arrows) and azimuthal velocity $\avg{u_{\phi}}{t}$ (colors) $(a,\,b)$ at height of thermal BL, $z=\delta_{\theta}\equiv H/(2\Nu)$ and $(c,\,d)$ at the mid-plane, $z=H/2$, with $\Ra=10^9$ and $1/\Ro=$ 0.5 $(a,\,c)$  and 10 $(b,\,d)$. 
Blue (pink) color indicates anticyclonic (cyclonic) motion. In $(d)$, locations $r=r_0$ of $\avg{u_{\phi}}{t}=0$ (solid line) and $r=r_{u_\phi^{\max}}$ of the maximum of $\avg{u_{\phi}}{t}$ (dashed line) are shown. 
}
\label{PIC2}
\end{figure}

\begin{figure*}
\includegraphics{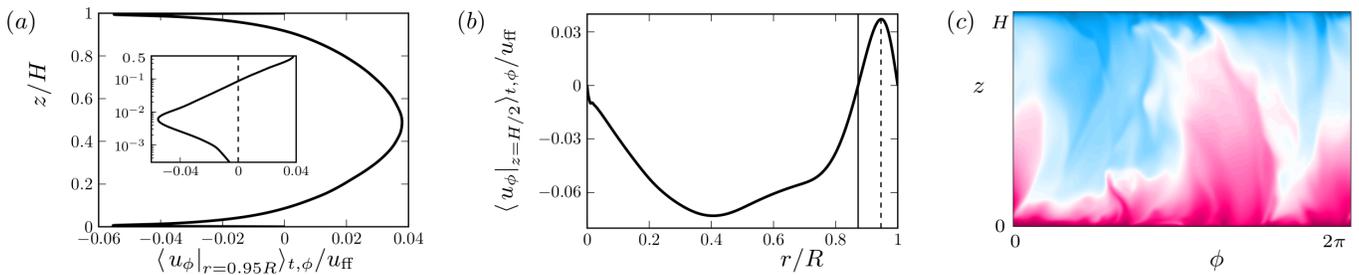}
\caption{
For $\Ra=10^9$, $1/\Ro=10$: $(a)$  $\langle u_\phi \rangle_{t,\phi}$ versus $z$ at $r=0.95R$. The inset shows the same data for $0\leq~z\leq~H/2$ in a log-plot. $(b)$  $\langle u_\phi \rangle_{t,\phi}$ versus $r$ at $z = H/2$; radial zero crossing $r=r_0$ (solid line) and radial maximum $r = r_{u_\phi^{\max}}$ (dashed line). $(c)$ Instantaneous thermal field at $r = r_{u_\phi^{\max}}$ versus $z$ and $\phi$.
}
\label{PIC3}
\end{figure*}

\begin{figure}
\includegraphics{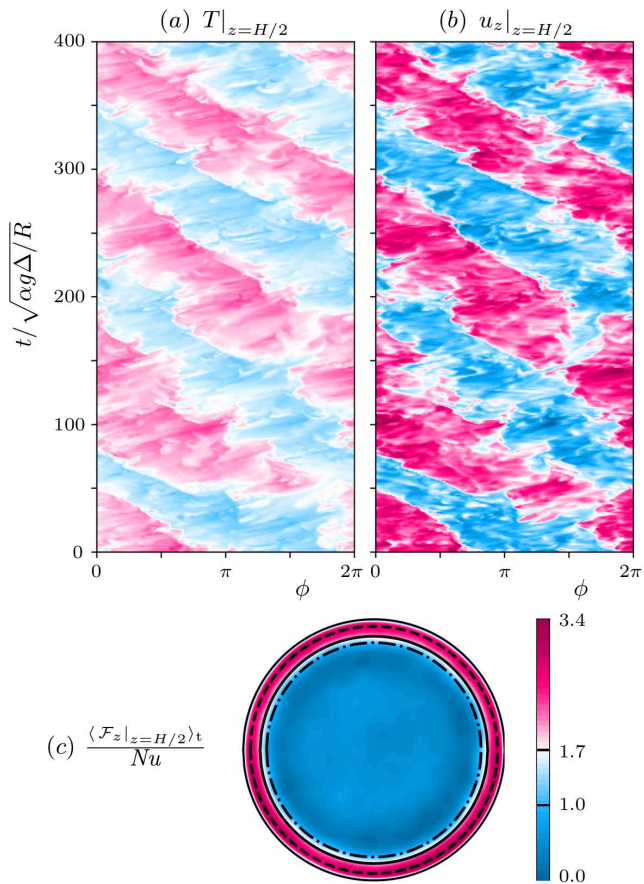}
\caption{
For $\Ra=10^9$ and $1/\Ro=10$: $(a,b)$ angle-time plots at $r=r_{u_\phi^{\max}}$, $z=H/2$ of $(a)$ $T$ and $(b)$ $u_z$; $(c)$ normalized time-averaged vertical heat flux $\avg{\mathcal{F}_z}{t}$ at $z=H/2$.
In $(c)$, location of $r$ where $\avg{\left.\mathcal{F}_z\right|_{z=H/2}}{t}=\Nu$  (dash-dotted line) and locations $r=r_0$ of $\avg{u_{\phi}}{t}=0$ (solid line) and $r=r_{u_\phi^{\max}}$ of the maximum of $\avg{u_{\phi}}{t}$ (dashed line) are shown. 
Color scale from blue (min values) to pink (max values) ranges $(a)$ between the top and bottom temperatures, $(b)$ in $[-u_\text{ff}/2,\,u_\text{ff}/2]$, $(c)$ from 0 to mid-plane magnitude of $\avg{\mathcal{F}_z}{t}$, which is $\approx3.4\Nu$.
}
\label{PIC4}
\end{figure}

\begin{figure*}
\includegraphics{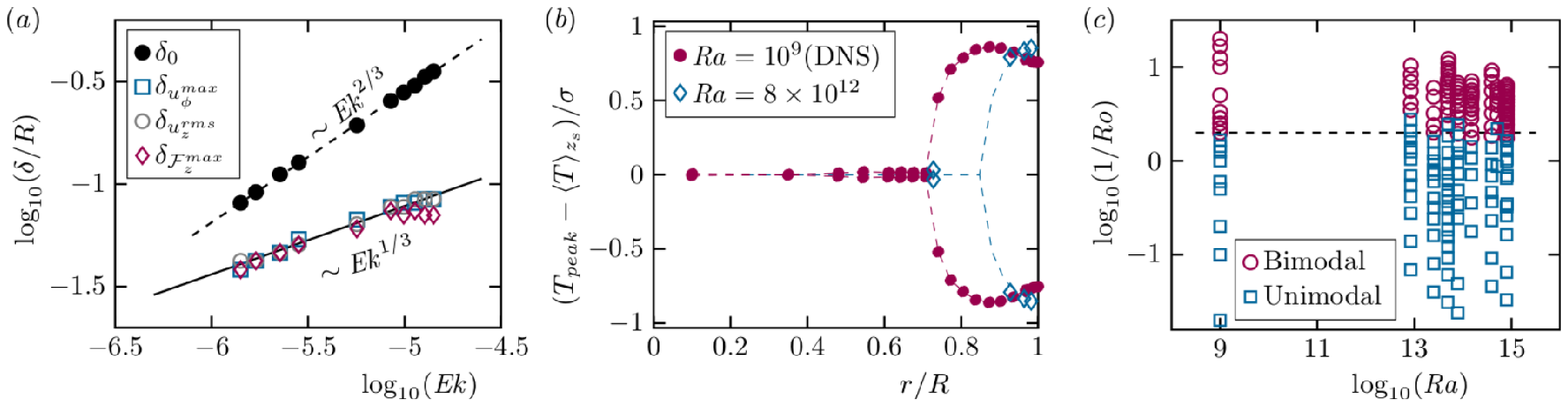}
\caption{
{\color{black}
$(a)$ Scaling of BZF widths $\delta_0$, $\delta_{u_\phi^{\max}}$, $\delta_{u_z^\text{rms}}$, and $\delta_{\mathcal{F}_z^{\max}}$ with $\Ek$ (DNS for $\Ra = 10^9$); $(b)$ Fitted peak values of bimodal PDF distributions (normalized by $\sigma$, standard deviation of $T$) at $z/H=1/2$ versus $r/R$: DNS ($\Ra=10^9$) and measurements ($\Ra=8\times10^{12}$), both for  $1/\Ro=10$;}
$(c)$ diagram of the bimodal and unimodal temperature distributions at $r=R$, according to our DNS ($\Ra = 10^9$) and experiments (larger $\Ra$) of rotating RBC for $\Pran\approx0.8$ and $\Gamma=1/2$. Critical inverse Rossby number equals $1/\Ro_c=2\pm1$ (shown with a dashed line).
}
\label{PIC5}
\end{figure*}

Turbulent fluid motion driven by buoyancy and influenced by rotation is a common phenomenon in nature and is important in many industrial applications. In the widely studied laboratory realization of turbulent convection, Rayleigh--B\'enard convection (RBC) \cite{Ahlers2009, Lohse2010}, a fluid is confined in a convection cell with a heated bottom, cooled top, and adiabatic vertical walls. For these conditions, a large scale circulation (LSC) arises from cooperative plume motion and is an important feature of turbulent RBC \cite{Ahlers2009}.  The addition of rotation about a vertical axis produces a different type of convection as  thermal plumes are transformed into thermal vortices, over some regions of parameter space heat transport is enhanced by Ekman pumping \cite{Rossby1969, Zhong1993, Liu1997, Kunnen2006, King2009, Zhong2009, Liu2009, Stellmach2014}, and statistical measures of vorticity and temperature fluctuations in the bulk are strongly influenced \cite{Julien1996, Hart2002, Vorobieff2002, Kunnen2008, Julien2012, Liu2011, Ding2019}.  A crucial aspect of rotation is to suppress, for sufficiently rapid rotation rates, the LSC of non-rotating convection  \cite{Hart2002, Vorobieff2002, Kunnen2008b, Zhong2010}, although the diameter-to-height aspect ratio $\Gamma = D/H$ appears to play some role in the nature of the suppression \cite{Weiss2011}.   

{\color{black} 
In RBC geometries with $1/2 \leq \Gamma \leq 2$, the LSC usually spans the cell in a roll-like circulation of size $H$.} 
For rotating convection, the intrinsic linear scale of separation of vortices is reduced with increasing rotation rate~\cite{Chandrasekhar1961, Julien2012a}, suggesting that one might reduce the geometric aspect ratio, i.e., $\Gamma < 1$ while maintaining a large ratio of lateral cell size to linear scale \cite{Liu1997}; such convection cells are being implemented in numerous new experiments \cite{Cheng2018}.  Thus, an important question about rotating convection in slender cylindrical cells is {\color{black} whether} there is a global circulation that substantially influences the internal state of the system and carries appreciable global heat transport.  Direct numerical simulations (DNS) of rotating convection~\cite{Kunnen2011} in cylindrical geometry with $\Gamma = 1$, inverse Rossby number $1/\Ro=2.78$, Rayleigh number $\Ra=10^9$ and Prandtl number $\Pran=6.4$ ($\Ro$, $\Ra$ and $\Pran$ defined below) revealed a cyclonic azimuthal velocity boundary-layer flow surrounding a core region of anti-cyclonic circulation and localized near the cylinder sidewall. The results were interpreted in the context of sidewall Stewartson layers driven  by active Ekman layers at the top and bottom of the cell \cite{Stewartson1957, Stewartson1966}.  

Here we show through DNS and experimental measurements for a cylindrical convection cell with $\Gamma=1/2$ at large $\Ra$ and for a range of rotation rates from slow to rapid {\color{black} that a wider (several times the Stewartson layer width) annular flow, denoted a boundary zonal flow (BZF), has profound effects on the overall flow structure and on the spatial distribution of heat flux. In particular, this cyclonic zonal flow surrounds an anti-cyclonic core. The BZF has alternating temperature sheets that produce bimodal temperature distributions for radial positions $r/R > 0.7$ and that contribute greatly to the overall heat transport; 60\% of heat transport are carried in the BZF.  Although the location of the azimuthally-averaged maximum cyclonic azimuthal velocity, the root-mean-square (rms) vertical velocity fluctuations, and the normalized vertical heat transport at the mid-plane are consistent with a linear description of a Stewartson-layer scaling \cite{Kunnen2011}, the dynamics of temperature, vertical velocity and heat transport in the BZF are more complex and interesting. The robustness of the BZF state as evidenced by its existence over 7 orders of magnitude in $\Ra$ in DNS and experiment and over a range $1/2\leq \Gamma \leq2$ and $0.1 \leq \Pran \leq 4.4$ (results to be presented elsewhere) suggests that this is an universal state of rotating convection that needs a physical understanding.}

The dimensionless control parameters in rotating RBC are the Rayleigh number $\Ra = \alpha g \Delta H^3/(\kappa \nu)$, Prandtl number $\Pran = \nu/\kappa$, cell aspect ratio $\Gamma$ and Rossby number $\Ro = {\sqrt{\alpha g\Delta H}}\;/\,({2 \Omega H})$ or, alternatively, Ekman number $\Ek = \nu/(2\Omega H^2)$.
Here $\alpha$ is isobaric thermal expansion coefficient, $\nu$ kinematic viscosity, $\kappa$ fluid thermal diffusivity, $g$  acceleration of gravity, $\Omega$ angular rotation rate, and $\Delta$ temperature difference between the hotter bottom and colder top plates. 
The main integral response parameters we consider is the Nusselt number $\Nu\equiv\langle\mathcal{F}_z\rangle_{t,V}$, where $\langle\cdot\rangle_{t,V}$ denotes the time- and volume-averaging and $\mathcal{F}_z\equiv(u_z(T-T_0)-\kappa\partial_zT)/(\kappa\Delta/H)$ is the normalized vertical heat flux with $u_z$ being the vertical component of the velocity and $T_0$ the average of the top and bottom temperatures. 

We present numerical and experimental results~\cite{DNSEXPlist} for rotating RBC in a $\Gamma=1/2$ cylindrical cell and $1/\Ro = 0$, 0.5 and 10. 
The DNS used the {\sc goldfish} code \cite{Kooij2018, Shishkina2015} with $\Pran = 0.8$ and $\Ra = 10^9$. 
The experiments used pressurized sulfur hexafluoride (SF$_6$) and were performed over a large parameter space in the High Pressure Convection Facility (HPCF, 2.24~m high) at the Max Planck Institute for Dynamics and Self-Organization in G\"ottingen~\cite{Ahlers2009c}. In the studied parameter range, the Oberbeck--Boussinesq approximation is valid \cite{Horn2014,Shishkina2016d, Weiss2018}, and the centrifugal force is negligible \cite{Zhong2009, Horn2015, Horn2018}.

We first consider the azimuthal variation of the temperature measured by thermal probes at or near the sidewall, a commonly used technique for parameterizing the LSC in RBC \cite{Cioni1997, Brown2005, Kunnen2008b, Weiss2011, Vogt2018b}. We measured experimentally and in corresponding DNS the temperature at 8 equidistantly spaced azimuthal locations of the sensors for each of 3 distances from the bottom plate: $z/H=1/4$, $1/2$ and $3/4$. 
The PDFs of the experimental data without rotation ($1/\Ro=0$, $Ra = 8 \times 10^{12}$) in Fig.~\ref{PIC1}a show a distribution with a single peak and slight asymmetry to hotter (colder) fluctuations for heights smaller (larger) than $z/H = 1/2$, whereas the PDFs for rapid rotation ($1/\Ro=10$, Fig.~\ref{PIC1}b), show a bimodal distribution that is well fit by the sum of two Gaussian distributions. The corresponding PDFs of the DNS data (at $\Ra = 10^9$) show the same qualitative transition from a single peak without rotation to a bimodal distribution in the rapidly rotating case with similar hot/cold asymmetry for different $z$ (Fig.~\ref{PIC1}c,~d).  To understand the nature of the emergence of a bimodal distribution near the radial boundary, we consider the DNS data in detail.

The LSC for non-rotating convection in cells with $1/2 \leq \Gamma \leq 2$ and at large $\Ra$ extends throughout the entire cell with a large roll-like circulation 
\cite{Shishkina2014}.  
With slow rotation, Coriolis forces induce anticyclonic motion close to the plates owing to the diverging flow between the LSC and the corner rolls. At the mid-plane, the LSC is tilted with a small inward radial velocity component that rotation spins up into cyclonic motion. These tendencies are illustrated for $1/\Ro = 0.5$ in Figs.~\ref{PIC2}a,~c, respectively, where streamlines of time-averaged velocity are overlaid on the field of azimuthal velocity. Fig.~\ref{PIC2}a shows fields evaluated at the thermal BL height $z = \delta_\theta\equiv H/(2 Nu)$,  demonstrating the dominant anticyclonic flow near the boundary. The situation is reversed at the mid-plane (Fig.~\ref{PIC2}c) where cyclonic motion extends over almost the entire cross sectional area.

For rapid rotation, viscous Ekman BLs near the plates induce anticyclonic circulation with radial outflow in  horizontal planes as in Fig.~\ref{PIC2}b. The outflow is balanced by the vertical velocity in an increasingly thin (with increasing $1/\Ro$) annular region near the sidewall where cyclonic vorticity is concentrated at the mid-plane, see Fig.~\ref{PIC2}d. The core region, on the other hand, is strongly anticyclonic owing to the Taylor-Proudman effect \cite{Proudman1916, Taylor1921} that tends to homogenize vertical motion. The circulation for a rotating flow in a finite cylindrical container consists of thin anticyclonic Ekman layers on top and bottom plates and compensating Stewartson layers along the sidewalls with up-flow from the bottom and down-flow from the top \cite{Kunnen2011, Kunnen2013}. This classical BL analysis was successfully applied to rotating convection \cite{Kunnen2011} for a $\Gamma = 1$ cylindrical cell with $\Pran = 6.4$ and $10^8 \leq \Ra \leq 10^9$ in both experiment and DNS. No evidence for a coherent large-scale circulation for rapid rotation was found in those studies.

For our conditions, $\Pran = 0.8$, $\Ra = 10^9$, and $1/\Ro = 10$, we compute the time- and azimuthal-average azimuthal velocity $\langle u_\phi \rangle_{t, \phi}$ (normalized by the free-fall velocity $u_\text{ff} = \sqrt{\alpha g \Delta/R}$) as a function of height $z$ for fixed $r = 0.95R$ and of radius $r$ at fixed $z = H/2$.
The height dependence of $\langle u_\phi \rangle_{t, \phi}$, Fig.~\ref{PIC3}a, shows an anticyclonic (negative) circulation close to the top and bottom plates and an increasingly cyclonic (positive) circulation with increasing (decreasing) $z$ from the bottom (top) plate.  The radial dependence, Fig.~\ref{PIC3}b, demonstrates the sharp localization of cyclonic motion near the sidewall as parameterized by the zero-crossing $r_{0}$ (solid line) and the maximum $r_{u_\phi^{\max}}$ (dashed line). Corresponding distances from the sidewall are $\delta_0 = R - r_{0}$ and $\delta_{u_\phi^{\max}}= R - r_{u_\phi^{\max}}$ where $\delta_{u_\phi^{\max}} \approx \delta_{u_z^\text{rms}}$ (based on  maximum of rms of $u_z$).
$\delta_{u_z^\text{rms}}$ was used to define the sidewall Stewartson layer thickness in rotating convection \cite{Kunnen2011}, and our results for $\langle u_\phi \rangle_t$ are consistent with that description.  What was absolutely {\it not} expected is the strong azimuthal variation of the instantaneous temperature $T$ shown in Fig.~\ref{PIC3}c, a feature that {\color{black} defines} the global flow circulation, namely the spatial distribution of the heat transport which is the origin of the bimodal temperature distributions seen in the experiments and DNS.

The strong variations in instantaneous temperature shown in Fig.~\ref{PIC3}c organize into anticyclonic traveling waves illustrated in the angle-time plot of $T$, Fig.~\ref{PIC4}a.  
The BZF height is order $H$, Fig.~\ref{PIC3}c, but is increasingly localized in the radial direction as the rotation rate increases ($\Ro$ and $\Ek$ decrease) so that $\delta_0/R \ll 1$. 
The azimuthal mode of $T$ is highly correlated with a corresponding mode of the vertical velocity, Fig.~\ref{PIC4}b, with a resulting coherent mode-1 ($m = 1$)  anticyclonic circulation in $\phi$ with a warm up-flow on one side of the cell balanced by a cool down-flow on the other side of the cell (for $\Gamma = 1, 2$, the dimensionless wave number $m/\Gamma = 2$ is independent of $\Gamma$, to be presented elsewhere).  The anticyclonic circulation is the speed of the anticyclonic horizontal BL, suggesting that the thermal wave is anchored at the horizontal BLs so that it travels against the cyclonic circulation near the sidewall. The coherence between $T$ and $u_z$ leads to localization of vertical heat flux near the sidewall shown in Fig.~\ref{PIC4}c where the heat flux within the annular area defined by $\delta_0$ is $\approx$60\% of the total heat flux.

We arrive at a compact description of the BZF.  The radial distances from the sidewall $\delta_{u_z^\text{rms}}$, $\delta_{\mathcal{F}_z^{\max}}$, and $\delta_{u_\phi^{\max}}$ of maxima of $u_z$-rms, heat flux $\mathcal{F}_z$, and $u_\phi$, respectively, scale as $Ek^{1/3}$, Fig.~\ref{PIC5}a, consistent with the expectations of Ekman/Stewartson BL theory \cite{Kunnen2011,Kunnen2013}.  On the other hand, the cyclonic zone width $\delta_0$ decreases more rapidly with $Ek$, i.e., as $Ek^{2/3}$ with a $Ra^{1/4}$ dependence (presented elsewhere).  Thus, the inner layer is consistent with Stewartson theory whereas the outer structure reflects the more complex character of interacting thermal and velocity fields.  The bimodal temperature distribution  is now explained by the alternating thermal field. We plot the radial dependence of the mean values of the bimodal distributions (the bimodal PDFs are well fit by the sum of two Gaussians) from the DNS for $\Ra=10^9$, $1/\Ro = 10$ in Fig.~\ref{PIC5}b.  
The unimodal distribution for small $r/R$ bifurcates sharply to a bimodal distribution for $r/R \approx 0.72$. 
{\color{black} The corresponding experimental measurements do not provide data at intermediate $r/R$, but are consistent (dashed curve) with a scaled BZF width based on the scaling $\Ra^{1/4}\,\Ek^{2/3}$. Finally, the transition value of $1/\Ro \approx 2$ from unimodal to bimodal distributions is roughly independent of $\Ra$ as indicated  in Fig.~\ref{PIC5}c.   

Our observations provide insight into experimental results for $\Gamma = 1/2$ in water with $\Pran = 4.38$~\cite{Weiss2011}, where the mode-1 LSC for non-rotating convection was reported to transition into a then unknown state. Our BZF is that unknown global mode.   
 We conclude that the BZF exists over a broad range of parameters $1/2 \leq \Gamma \leq 2$, $0.1 \leq \Pran \leq 4.4$, and $10^8 \leq \Ra < 10^{15}$ (details to be published elsewhere).
Here we presented details for $\Pran=0.8$ and $\Gamma = 1/2$ and for $\Ra$ spanning seven orders of magnitude \cite{DNSEXPlist}. 
A fully quantitative  understanding remains a challenge for the future. 
}

\acknowledgements
The authors acknowledge support from the Deutsche Forschungsgemeinschaft (DFG) through the Collaborative Research Centre SFB~963 "Astrophysical Flow Instabilities and Turbulence" and research grants Sh405/4-1, Sh405/4-2, Sh405/8-1, Ho5890/1-1 and  We5011/3-1, from the LDRD program at Los Alamos National Laboratory and by the Leibniz Supercomputing Centre (LRZ).


%

\end{document}